\documentclass{aa}
\usepackage{natbib}
\bibpunct{(}{)}{;}{a}{}{,}
\usepackage{psfig}

\def\gsim{\;\lower4pt\hbox{${\buildrel\displaystyle >\over\sim}$}\,}
\def\lsim{\;\lower4pt\hbox{${\buildrel\displaystyle <\over\sim}$}\,}

\def \xmm {{\em XMM-Newton}}
\def \src {IC443}

\def \nh {N${\rm _H}$}

\def \hcm {\hbox {\ifmmode $ atom cm$^{-2}\else atom cm$^{-2}$\fi}}
\def \arcmin {\hbox{$^\prime$}}
\def \arcsec {\hbox{$^{\prime\prime}$}}

\def\approxgt{\mathrel{\hbox{\rlap{\lower.55ex \hbox {$\sim$}}
        \kern-.3em \raise.4ex \hbox{$>$}}}}
\def\approxlt{\mathrel{\hbox{\rlap{\lower.55ex \hbox {$\sim$}}
        \kern-.3em \raise.4ex \hbox{$<$}}}}

\newcommand {\Msun}{M_\odot}

\def\lsim{\;\raise0.3ex\hbox{$<$\kern-0.75em\raise-1.1ex\hbox{$\sim$}}\;}
\def\gsim{\;\raise0.3ex\hbox{$>$\kern-0.75em\raise-1.1ex\hbox{$\sim$}}\;}
\def \gr{$\gamma$-ray }

\def\beq{\begin{equation}}
\def\enq{\end{equation}}
\def\begar{\begin{eqnarray}}
\def\endar{\end{eqnarray}}
\def\mathnew{\mathsurround=0pt}
\def\simov#1#2{\lower .5pt\vbox{\baselineskip0pt \lineskip-.5pt
        \ialign{$\mathnew#1\hfil##\hfil$\crcr#2\crcr\sim\crcr}}}

\def\cmc{\rm ~cm^{-3}}

\def\kms{\rm ~km~s^{-1}}

\def\enf{\rm ~erg~cm^{-2}~s^{-1}}

\def\etal{{ et al. }}

\def \chan {{\it Chandra}}
\def \xmm {{\it XMM-Newton}}
\def \src {IC443}

\begin{document}


\title{XMM-Newton study of hard X-ray sources in IC443}


\author{F. Bocchino\inst{1}
\and A.M. Bykov\inst{2}
}

\institute{INAF, Osservatorio Astronomico di Palermo, Piazza del Parlamento 1,
       90134 Palermo, Italy
\and
       A.F. Ioffe Institute for Physics and Technology,
          St. Petersburg, 194021, Russia
      }

\offprints{e-mail: bocchino$@$astropa.unipa.it}

\date{Received 20 Aug 2002 / Accepted 16 Dic 2002}


\abstract{We present \xmm\ observations of hard X-ray emission from the
field of IC443, a supernova remnant interacting with a molecular
cloud.  The hard emission from the field is dominated by 12 isolated
sources having 2--10 keV flux $\ga 10^{-14}$ $\enf$. Only
a fraction of the sources are expected to be extragalactic or
stars on statistical
grounds, while the others may be associated with the remnant.  We have
analyzed near-infrared K band and also DSS optical data for all of the
detected sources, finding that six X-ray sources are located in a
relatively small $15^\prime\times 15^\prime$ region where there is
strong 2.2 $\mu {\rm m}$ infrared emission, indicating interaction with a
molecular cloud.  The source 1SAX J0618.0+2227, the brightest in this
region (excluding the plerion), is resolved with \xmm\ into two
sources, one of which is extended and has a hard power law spectrum
photon index $\lsim$ 1.5) and shows some indications of spectral line
signatures (e.g. Si), while the other is point-like and has a
featureless spectrum of steeper photon index $\sim$ 2.2.  Possible
interpretations of some of the discrete sources in terms of interaction
between the SNR and the molecular cloud are discussed.

 }

\maketitle

\section{Introduction}
The middle-aged SNR IC443 (G189.1+3.0) appears in multiwavelength
images as an extended ($\sim$ 45 arcmin size) source of both
nonthermal and thermal continuum emissions with rich line spectra
from radio to X-rays. The extended component of the thermal soft
X-ray emission from IC 443 has been observed with the {\it
Einstein} satellite and the {\it HEAO 1 A2} experiment
(\citealt{pss88}), {\it ROSAT}
(\citealt{aa94}) and {\it ASCA} (\citealt{kpg97}; \citealt{kon02}).
On the other hand, the hard X-ray emission from IC443 in the 2--10
keV band, detected by \citet{wah92} using data of the {\it Ginga}
satellite, has been studied in detail only recently.
\citet{kpg97}, using the ASCA X-ray observatory, mapped for the
first time the hard X-ray emission of IC443, concluding that most
of the 2--10 keV photons came from an isolated emitting feature
and from the southeast elongated ridge of hard emission. They also
suggested that synchrotron emission of $\gsim$ 10 TeV regime
electrons provides the explanation for these observed features.
BeppoSAX observations in the hard X-ray band resolved the two ASCA
sources into two compact sources of hard emission up to 10 keV
(1SAX J0617.1+2221 and 1SAX J0618.0+2227) and a harder component
up to 100 keV, indicating that a synchroton origin is unlikely, and
suggesting electron acceleration by a slow shock in a molecular
cloud (\citealt{bb00}).  While recent observations of 1SAX J0617.1+2221
with \chan\ (\citealt{ocw01}) and \xmm\ (\citealt{bb01}) have
established the plerionic nature of this source, the debate is
still open about the nature of 1SAX J0618.0+2227.

The morphology of IC443 at high X-ray energies is therefore
expected to be very interesting. It is important to note that
IC443 is associated with the EGRET source 3EG J0617+2238
(\citealt{ehk96}), but the relation between the remnant and the
$\gamma$-ray source is not yet clear. In fact, the 95\% error
circle of the most recent-derived sky position of 3EG J0617+2238,
reported by \citet{hbb99}, does not include the plerion (see Fig.
\ref{softpn}), and the other compact X-ray sources, even if the
extrapolation of the X-ray flux of 1SAX J0617.1+2221 and 1SAX
J0618.0+2227 to the EGRET regime indicates that there could be a
direct relation (\citealt{bb00}).  High-resolution hard-X-ray
studies are therefore of great importance to understand this
complicated object, also in the light of the recent discovery by
\citet{uta02} of very similar features in the hard X-ray emission
of the SNR $\gamma$\  Cygni observed by the ASCA X-ray satellite.
Compact sources with hard emission spectra in both IC 443
and in $\gamma$ Cygni could be of similar nature e.g. shocked
molecular clumps as was predicted by the model of Bykov \etal
(2000).

Another interesting class of hard X-ray  sources related to the SNR
could be fast moving fragments of metal-rich ejecta interacting
with the ambient medium.
The isolated X-ray emitting ejecta fragments were
discovered in the middle-aged Vela SNR (\citealt{aet95}) and also in
type Ia Tycho SNR (e.g. \citealt{dsa01}). A model of X-ray line
emission from fast moving ejecta fragments predicts that the fragments
interacting with molecular clouds are the brightest and they could be
detected from a few kpc distances (\citealt{byk02}).

In this paper, we are not discussing the IC443 plerion, which was
already described by \citet{ocw01} and \citet{bb01}, while we will
address in detail 1SAX J0618.0+2227, a source closely correlated
with the molecular line emission region (\citealt{bb00}).
Moreover, the combination of high sensitivity of the PN and MOS
cameras aboard \xmm\ to 2--10 keV photons with a few arcseconds
angular resolution has allowed us to study the hard X-ray emission
from the whole extended supernova remnant IC443 in great detail,
thus discovering an additional extended component and a new set of
weaker compact sources. Both were undetected in previous BeppoSAX
and ASCA observations. In the present paper using the Cal/PV
dedicated pointings on IC443 we present the counting and analysis
of all the hard sources of X-ray emission within the boundary of
the IC443 remnant with 2--10 keV flux $\gsim 5\times 10^{-14}$
$\enf$, while a study of the diffuse hard X-ray emission of this
remnant will be discussed elsewhere.

\section{Observations and Data analysis}


\src\  has been observed as part of the Cal/PV phase of the \xmm\
satellite (\citealt{jla01}). In this paper, we have used the observations
listed in Table \ref{obs}.  Data from the two MOS (\citealt{taa01})
cameras and the PN (\citealt{sbd01}) camera have been used.  MOS and PN
cameras are CCD arrays which collect X-ray photons between 0.1 and 15
keV and have a field of view of $30^\prime$. The pixel size is
$1.1\arcsec$ and $4.1\arcsec$ for MOS and PN respectively, while the
mirror PSF width is $6\arcsec - 15\arcsec$ FWHM--HEW. The data have
been acquired with the medium filter and in full image mode, and
therefore the temporal resolution is low, 2.5 s and 73 ms for MOS and
PN, respectively. The worse spatial resolution of the PN is compensated
by its greater sensitivity, on the average 20-30\% more than the
combined two MOS.

\begin{table}
\caption{Cal/PV Observations of \src}
\label{obs}
\medskip
\centering\begin{minipage}{8.7cm}
\begin{tabular}{lcccc} \hline
Obs. & Coord (J2000) & T$_{us}$/T$_{s}$\footnote{Unscreened and screened exposure times} & Date \\
     &               & ksec &               \\ \hline

ptg1 & $6^{\rm h}17^{\rm m}24.2^{\rm s}$ $+22^{\rm d}43^{\rm m}13^{\rm s}$ & 23/10 & 26 Sep 2000 \\
ptg2\footnote{Two pointings combined}
     & $6^{\rm h}16^{\rm m}11.2^{\rm s}$ $+22^{\rm d}43^{\rm m}29^{\rm s}$ & 30/21 & 25 Sep 2000 \\
ptg3$^b$
     & $6^{\rm h}17^{\rm m}24.3^{\rm s}$ $+22^{\rm d}26^{\rm m}43^{\rm s}$ & 32/24 & 27 Sep 2000 \\
ptg4 & $6^{\rm h}16^{\rm m}11.2^{\rm s}$ $+22^{\rm d}19^{\rm m}29^{\rm s}$ & 30/23 & 28 Sep 2000 \\
\hline
\end{tabular}
\end{minipage}
\end{table}

The Standard Analysis System (SAS) software we have used (version
5.3) takes care of most of the required event
screening. However, we have further screened the data to eliminate some
residual hot pixels and occasional background enhancement due to
intense flux of soft protons in the magnetosphere. In particular, we
have extracted the background light-curve at energy $> 10$ keV and we
have identified time intervals of unusually high count rates (typically
more than 25 cnt s$^{-1}$) and removed them from subsequent analysis.
Moreover we have also selected events with values of the PATTERN
keywords between 0 and 12 for MOS and between 0 and 4 for PN.  The
exposure time of the screened observations is also reported in Table
\ref{obs}. In the case of $ptg2$ and $ptg3$, which are splitted into two
subpointings each, we have merged the event files of the subpointings
before continuing with the analysis.

\subsection{Image mosaic}

\begin{figure*}
  \centerline{\psfig{file=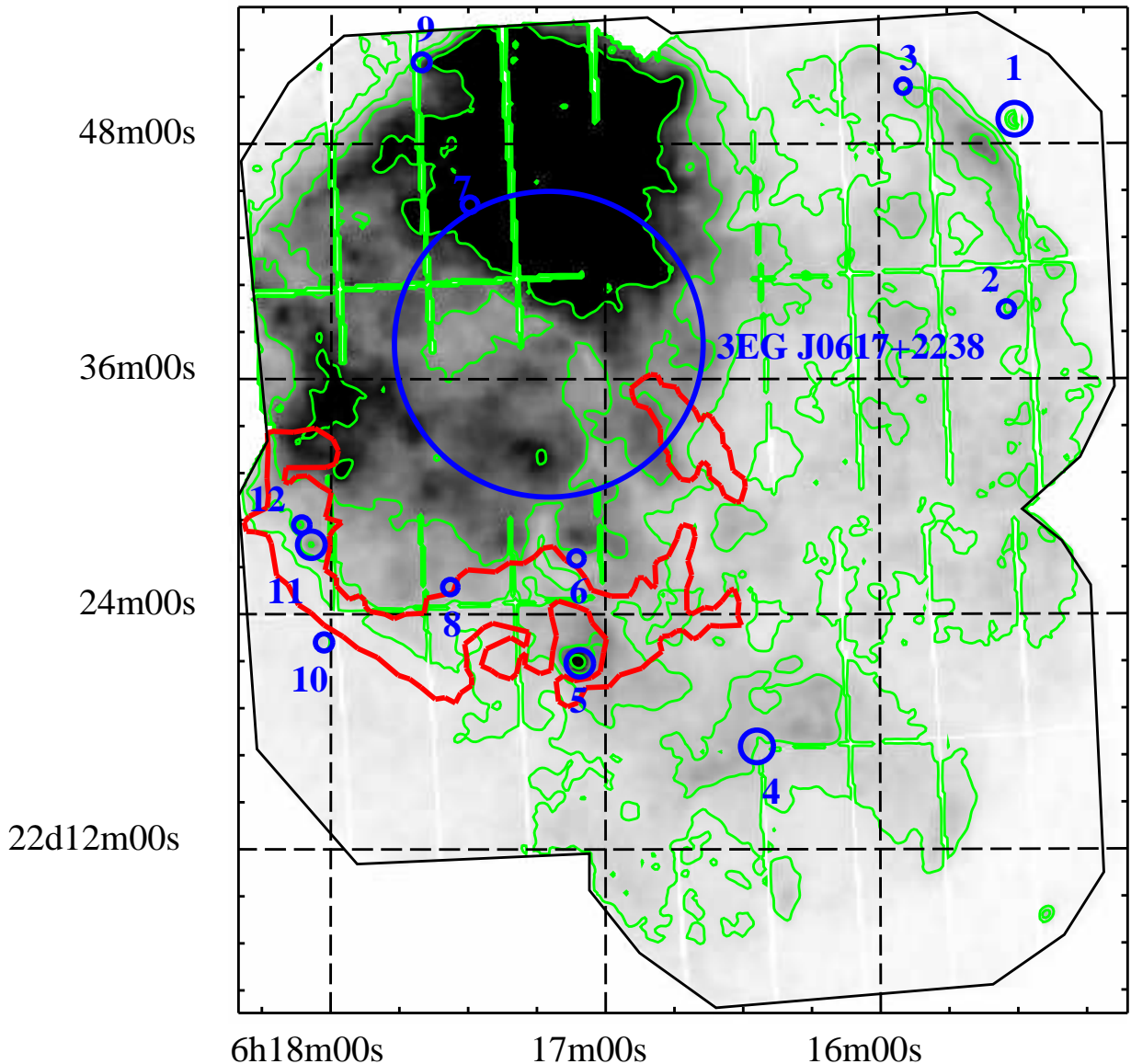,width=17.0cm}}
  \caption{PN mosaic in the soft (0.5--2 keV) energy band, with surface
  brightness contours at $5\times 10^{-4}$, 0.001, 0.002, and 0.004 cnt
  s$^{-1}$ pix$^{-1}$ ($8\arcsec\times 8\arcsec$ pixel). We have
  overlaid in bold style the 12 sources detected, the 95\% error circle of the
  position of the EGRET source 3EG J0617+2238, and a contour from the
  H$_2$ 1--0 $S(1)$ line emission map at 2.122 $\mu {\rm m}$ from
  \protect\citet{bhh90}, the emission line which is produced by shocked
  molecular hydrogen according to \protect\citet{bgb88}.}

  \label{softpn}
\end{figure*}

For each of the pointings listed in Table \ref{obs} we have extracted
PN images with 8\arcsec\  pixel size and exposure images at the same
resolution.  We have defined two energy bands in which to extract
images, namely 0.5--2 keV (hereafter soft band)
and 3--10 keV (hard band). The images in each band (and the exposure
maps) have been aligned using the reference pointings directions. The
resultant count image has been divided by the exposure maps mosaic to
correct for exposure and vignetting effect. Fig. \ref{softpn} shows the
0.5--2 keV mosaic of the PN pointings. Most of the emission in this
band is of thermal origin, as established by \citet{aa94} based on
ROSAT data.

\begin{figure*}[!ht]
  \centerline{\psfig{file=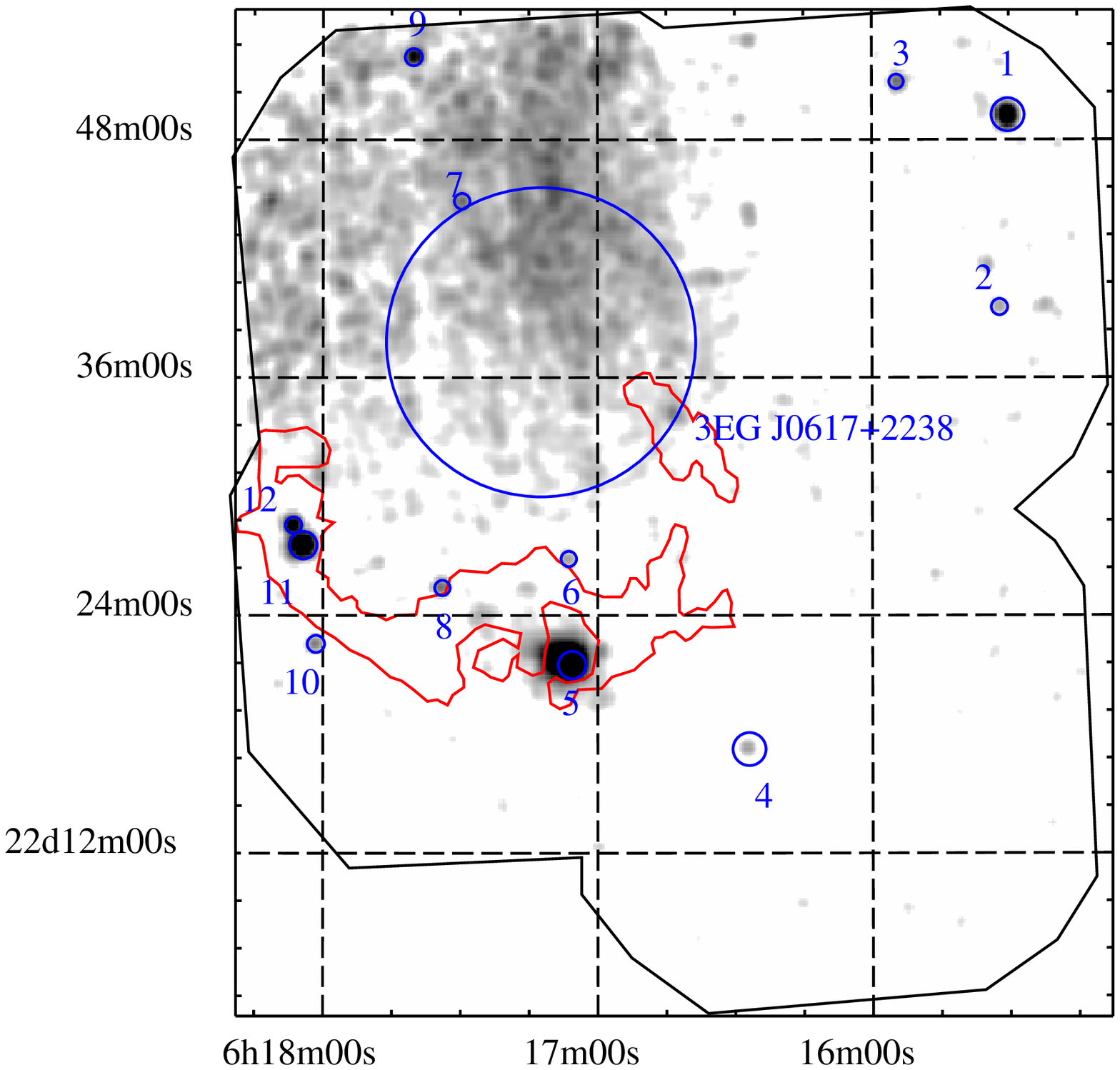,width=17.0cm}}
  \caption{PN Mosaic in the 3--10 keV energy band. Same overlays as for
  Fig. \protect\ref{softpn}. Src 5 is 1SAX J0617.1+2221, the plerion
  nebula studied in \protect\citet{ocw01} and \protect\citet{bb01},
  while Src 11 and 12 correspond to 1SAX J0618+2227
  (\protect\citealt{bb00}).}

  \label{hardpn}
\end{figure*}

Fig. \ref{hardpn} reports the 3--10 keV PN mosaics of IC443. Most of
the thermal emission associated to IC443 is not present in this band,
and therefore this band is very well suited for the study of hard compact
X-ray sources in the vicinity of this SNR.

\subsection{Compact sources}

\subsubsection{Source detection}

We have used the {\sc ewavelet} SAS task (v2.4) to detect sources in
the PN images.  This task uses the mexican hat wavelet algorithm to
detect both point and extended sources provided that the source extent
is not too large ($<1/8$th the image size). A review of detection
algorithms for \xmm, including {\sc ewavelet} is presented by
\citet{vpg01}, while further information about this task can be found
in the SAS User's Guide.  \citet{vpg01} have tested {\sc ewavelet} with
simulations of both point-like and extended sources, finding that the
algorithm performs very well for point-like sources, whereas it tends to
miss more extended sources than other algorithms (e.g. {\sc emldetect}).
Since we are mostly interested in compact hard X-ray sources, we have
discarded sources detected at large wavelet scales ($>6$ pixel), and
therefore this limitation does not affect our results. A crucial
parameter for source detection is the detection threshold;
\citet{vpg01} show that {\sc ewavelet} finds a number of false
detections of 10\% of the number of input sources in an observation of 10
ks with cosmic background, and using a threshold of $4\sigma$. To
further reduce the number of spurious detections, we have adopted a
threshold of $5\sigma$.

We have run the algorithm in the soft (0.5--2 keV) and hard (3--10 keV)
energy band, using wavelet scales in the range 2--8 pixels. We have
detected 88 sources in the soft band, and 12 in the 3--10 keV band.  We
therefore expect less than one spurious source in the hard band.  The
soft-band list is dominated by sources representing fluctuations in the
thermal soft X-ray emission of this remnant. For this reason, we have
reported in Table \ref{det} only the 3--10 keV band detections along
with their soft counterpart if any.  Figs. \ref{softpn} and \ref{hardpn}
also show the positions of the detected sources.

Table \ref{det} reports the source position, the wavelet correlation
coefficients which are a measure of source detection significance, the
wavelet scale of detection and the extension parameter, which is
equivalent to a FWHM in the approximation of a Gaussian Point Spread
Function. Simulations with point sources only have shown that this
parameter is $<20\arcsec$ in 97\% of the cases, and therefore indicates
reliably if a source is point-like or not.

\begin{table*}
\caption{PN detected sources.}
\label{det}
\medskip
\centering\begin{minipage}{13.0cm}
\begin{tabular}{rccrrrrrc}
\hline\noalign{\smallskip} N & RA & Dec & WCORR\footnote{The
maximum correlation coefficient is a measure of source
significance.} & FWHM & cts s$^{-1}$\footnote{The count rate is
given in the 2--10 keV band for comparison with other published
results. The flux was corrected for the PSF following the
suggestion of \citet{vpg01}. $10^{-3}$ cnt s$^{-1}$ is
corresponding to $7.8\times 10^{-15}$ $\enf$ adopting a power-law
emission model with $\gamma = 2$ and interstellar absorption
\nh$=7\times 10^{21}$ cm$^{-2}$.}
 & T$_{\rm exp}$ & Note\footnote{``S" marks sources with a counterpart in the soft band.}  \\
     & \multicolumn{2}{c}{J2000} & & arcsec & $\times 10^{-3}$ & sec \\
\noalign{\smallskip\hrule\smallskip}

1 &  6 15 30.1     &     22 49 16.7 & 5.8 & 12.0 $\pm$ 7.2 &  36.7 $\pm$  3.6 & 7107      & S \\
2 &  6 15 32.1     &     22 39 33.0 & 0.7 & 11.9 $\pm$ 2.8 &   6.3 $\pm$  1.3 & 9335 \\
3 &  6 15 54.6     &     22 50 55.6 & 1.3 & 11.3 $\pm$ 1.8 &   5.7 $\pm$  1.2 & 9994 \\
4 &  6 16 27.1     &     22 17 18.2 & 1.5 & 10.5 $\pm$ 1.4 &   2.8 $\pm$  0.6 & 20441     &  \\
5 &  6 17 05.5     &     22 21 29.7 & 11.0& 20.5 $\pm$ 1.2 &  65.8 $\pm$  4.4 & 16301 & S, Plerion \\
6 &  6 17 06.3     &     22 26 51.1 & 0.5 & 12.5 $\pm$ 3.8 &   4.8 $\pm$  0.9 & 18356 \\
7 &  6 17 29.6     &     22 44 53.9 & 0.7 & 11.0 $\pm$ 2.6 &   7.0 $\pm$  1.5 & 9398 & S \\
8 &  6 17 33.9     &     22 25 23.4 & 2.1 & 11.7 $\pm$ 1.3 &   5.8 $\pm$  0.9 & 22860 \\
9 &  6 17 40.2     &     22 52 10.2 & 0.2 & 53.0 $\pm$ 22.0&  13.1 $\pm$  2.3 & 4430 \\
10 &  6 18 01.5     &     22 22 33.7 & 1.3 & 11.3 $\pm$ 1.7 &  7.2 $\pm$  1.2 & 13382 & S \\
11 &  6 18 04.3     &     22 27 32.8 & 2.9 & 20.6 $\pm$ 2.2 & 24.7 $\pm$  3.2 & 12463 & S \\
12 &  6 18 06.4     &     22 28 32.8 & 3.3 & 12.2 $\pm$ 1.0 & 23.4 $\pm$  2.3 & 11471 & S \\

\noalign{\smallskip}
\hline
\end{tabular}
\end{minipage}
\end{table*}

\subsubsection{Sources identification}

In order to investigate the nature of the hard X-ray sources in IC443,
we have carried out an extensive set of cross correlations with
existing catalogues and digital data. In particular, we have used all
the catalogues present in the W3Browse facility of Goddard Space Flight
Center of NASA, and in the SIMBAD facility of the Centre de Donn\`ees
astronomiques de Strasbourg.  We have used a standard search radius of
2\arcmin\  for all the catalogues except astrometric optical catalogues
such as GSC, USNO and Tycho, were we have used 10\arcsec. The list of
identification is reported in Table \ref{ident}.

\begin{table*}
\caption{Identification of PN detected sources with catalogues.}
\label{ident}
\medskip
\centering\begin{minipage}{17cm}
\begin{tabular}{rcccrc}
\hline\noalign{\smallskip}
N & Source name & Name & Catalog & Offset & Remarks \\
     &          &  &  & arcsec \\
\noalign{\smallskip\hrule\smallskip}

1 &  XMMU J061530.1+224917 & U061530.75+224910.6 & USNO & 9 & B=18.8,R=15.8 \\
2 &  XMMU J061532.1+223933 & \\
3 &  XMMU J061554.6+225055 & 06128+2250 & IRASPSC & 84 & $f_{12,25,60,100}=3.1,0.9,<1.1,<8.2$ Jy\\
4 &  XMMU J061627.1+221718 & \\
5 &  XMMU J061705.5+222129 & U061705.98+222120.0 & USNO & 11 & B=16.7,R=15.9 \\
  &                         & 1WGA J0617.1+2221 & WGACAT & 6 \\
  &                         & H061706.08+222131.3 & ROSHRI & 6 \\
  &                         & 1SAX J0617.1+2221 &  &  & Src A in \citet{bb00} \\
6 &  XMMU J061706.3+222651 & \\
7 &  XMMU J061729.6+224453 & U061729.57+224452.8 & USNO & 1 & B=15.9,R=15.6 \\
8 &  XMMU J061733.9+222523 & TXS 0614+224 & TEXAS\footnote{The Texas
Survey of Radio Sources at 365 MHz of \citet{dbb96}.}
 & 34 & PosErr $<1\arcsec$ \\
9 &  XMMU J061740.2+225210 & \\
10 &  XMMU J061801.5+222233 & H061801.60+222229.8 & ROSHRI & 4 & \\
   &                        & RICH88 & DIXON\footnote{The Master List of Radio Sources (Version 43) of \citet{dix70}.} & 74 \\
11 &  XMMU J061804.3+222732 & 1SAX J0618.0+2227$^c$ &  & & Src B in \citet{bb00} \\
12 &  XMMU J061806.4+222832 & 1SAX J0618.0+2227\footnote{The BeppoSAX source 1SAX J0618.0+2227 is resolved in two sources by our XMM-Newton observation.}  &  & & Src B in \citet{bb00} \\

\noalign{\smallskip}
\hline
\end{tabular}
\end{minipage}
\end{table*}

\begin{figure*}
  \centerline{\psfig{file=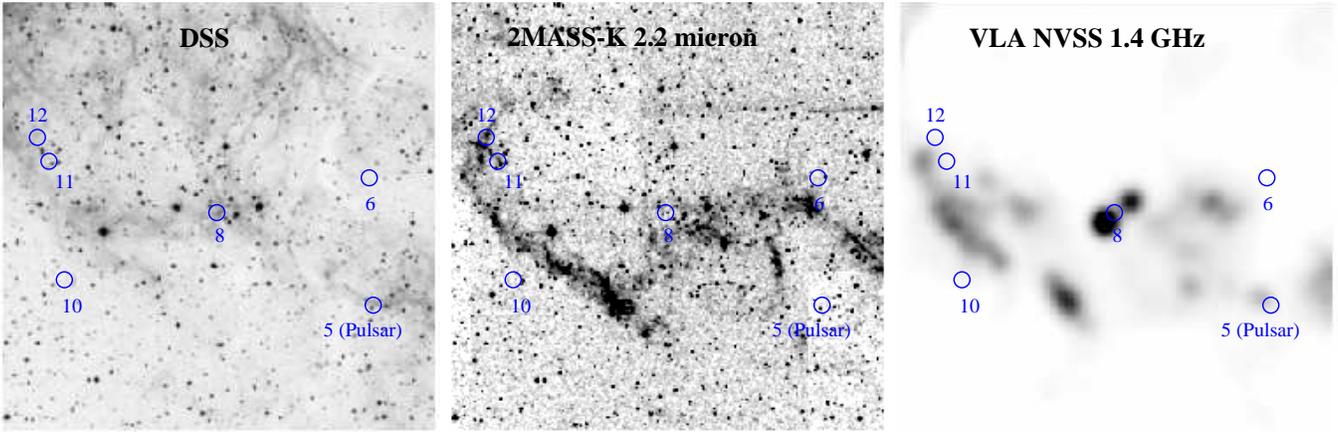,width=18.0cm,angle=-90}}
  \caption{Infrared/optical/radio images of the putative region of
  interaction with a molecular cloud.  {\em Left:} optical DSS image.
  {\em Center:} 2MASS K$_{\rm s}$ (2.32 $\mu {\rm m}$) band image.  {\em Right} VLA
  NVSS radio survey at 1.4 GHz.  All images are 18\arcmin\  size and we
  have superimposed the position of hard X-ray sources using a circle of
  $20^{\prime\prime}$ radius.}

  \label{cpart}
\end{figure*}

To verify if the hard X-ray sources are due to the interaction
between IC443 and the molecular cloud which lies in foreground
(\citealt{bgb88}, \citealt{rjc01}), as suggested by \citet{bb00},
we have exploited the high spatial resolution of the PN data for a
comparison with high-resolution optical and infrared data. We have
used the Digital Sky Survey and the near-infrared data of the Two
Micron All Sky Survey (2MASS, \citealt{cut98}), which has a
resolution of $\sim 3\arcsec$. In particular, we have considered
the K$_{\rm s}$ (2.32 $\mu {\rm m}$) band of 2MASS, for which most of the emission
can be explained by molecular H$_2$ rovibrational lines
(\citealt{rjc01}), a classical indicator of a shocked molecular
cloud. Fig.  \ref{cpart} shows the correspondence between the
infrared, optical and X-ray emission of the hard X-ray sources
located in the region of interaction between the SNR and the
molecular cloud.

\subsubsection{Hardness ratios of detected sources}

\begin{table}
\caption{Definition of hardness ratios.}
\label{defhr}
\medskip
\centering\begin{minipage}{8.5cm}
\begin{tabular}{lcc} \hline
Name & S$_{\rm i}$ & H$_{\rm i}$ \\
\hline
HR$_1$  & 0.2--0.5 & 0.5--2.0 \\
HR$_2$  & 0.5--2.0 & 2.0--4.5 \\
HR$_3$  & 2.0--4.5 & 4.5--10.0 \\
\noalign{\smallskip}
\hline
\end{tabular}
\end{minipage}
\end{table}

\begin{figure}
  \centerline{\psfig{file=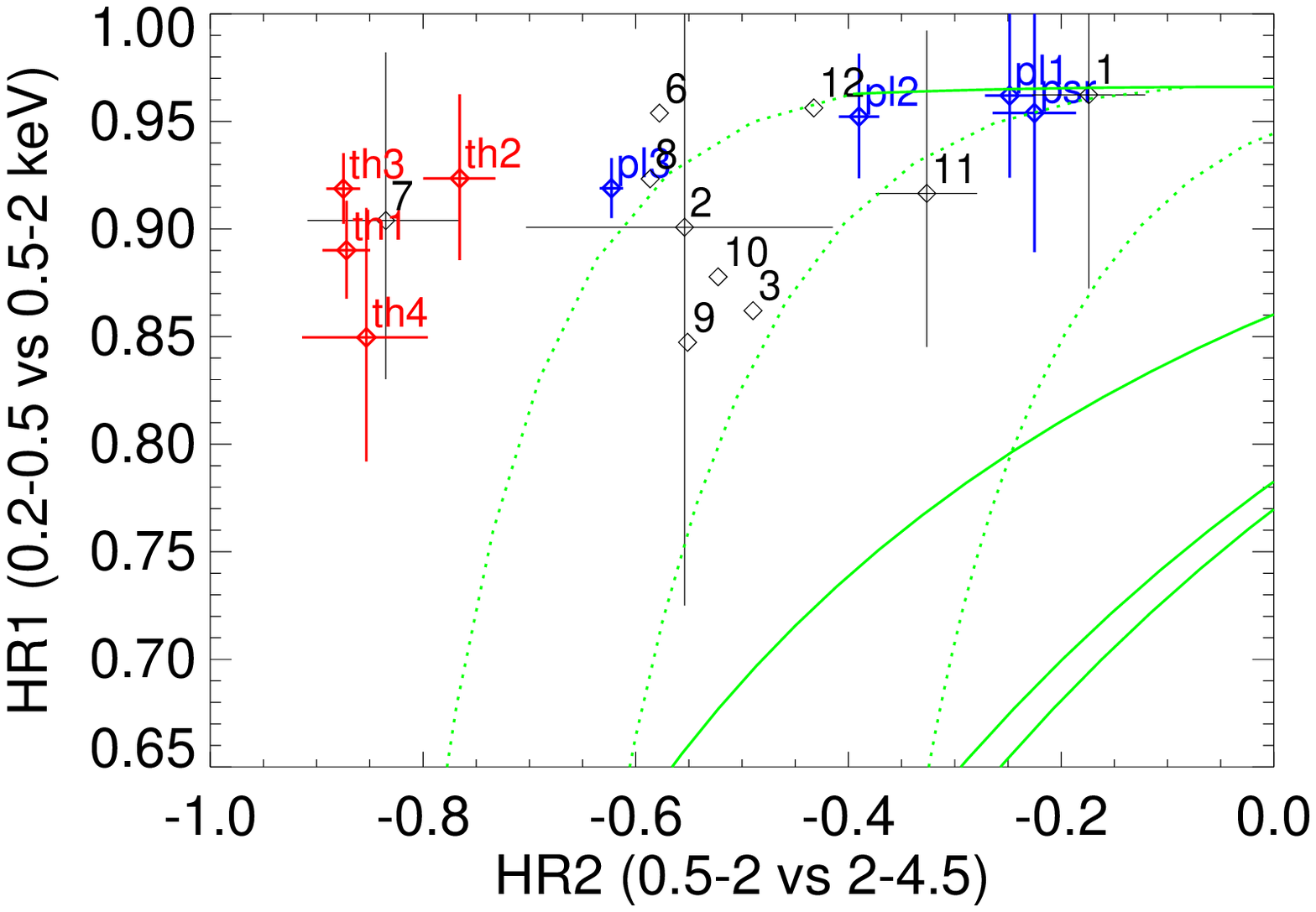,width=9.5cm}}
  \centerline{\psfig{file=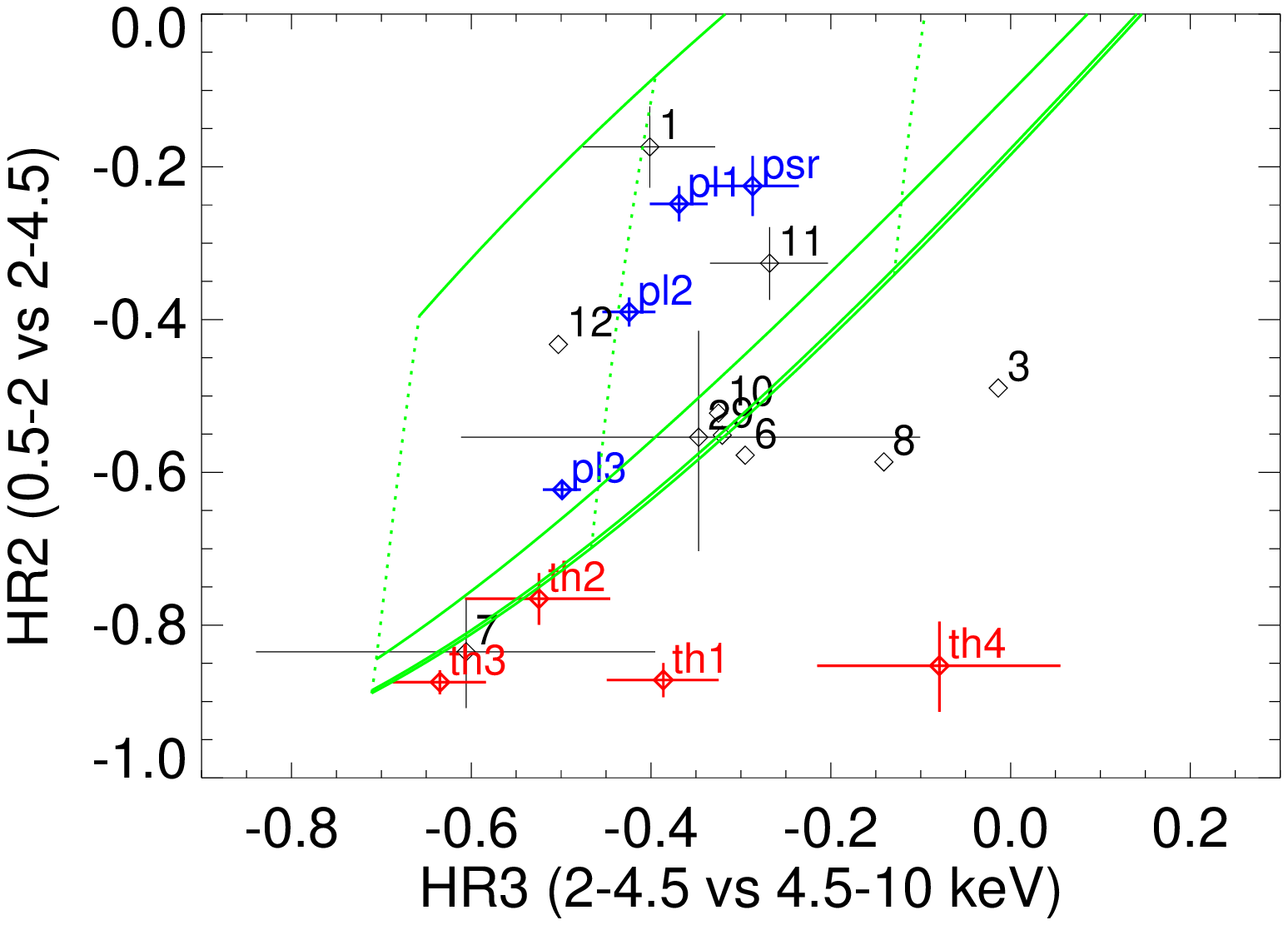,width=9.5cm}}
  \caption{Hardness ratios of the detected sources (1-12). We have also
  plotted for comparison the HRs of a sample of thermal (some knot of
  the shell diffuse emission, $th1 - th4$) and non-thermal (the plerion
  nebula, $pl1 - psr$) sources. The expected HR values for a power-law
  model are also shown as solid lines (constant \nh
  $=10^{20},10^{21},10^{22},10^{23}$ cm$^{-2}$, from bottom to top)
  and dotted lines (constant $\gamma=1,2,3$ from right to left). Some
  representative error bars are shown.}

  \label{hr}
\end{figure}

In order to characterize the spectral properties of the detected
source, we have computed three different and statistically
independent hardness ratios (HRs) of the form

\begin{equation}
HR_i = \frac{H_i-S_i}{H_i+S_i}; ~~~~~ i=1,2,3
\end{equation}

where the spectral bands are defined in Table \ref{defhr}. The
definition is the same as adopted by \citet{haa01} to
classify the sources detected in a deep XMM-Newton observation of
the Lockman hole. The hardness ratios of the detected sources are
displayed in Fig. \ref{hr}, in which we have also reported, for
comparison, the HRs of a sample of regions of the thermal shell
(Th1 - Th4), and the four regions of the plerion nebula defined in
\citet{bb01}. The latter can be described by a pure non-thermal
power law spectrum with a slope steepening toward the outer
regions (i.e. going from the core to the ``pl3" region, see Table 2 of
\citealt{bb01}).

Most the sources, except source number 7, clearly have a non-thermal
origin, and are located in a well-defined region of the diagram, near
``pl3", i.e. the most external plerion region which has a power-law
spectrum with a photon index of $2.3$.  Source 1, 11 and 12 are located in
a different group, near the regions ``psr" and ``pl1", thus implying an
intrinsically harder and/or more absorbed spectrum.

\subsubsection{Spectral analysis of the sources}

We have chosen to extract the spectra of the brightest sources which
correlate with the infrared filaments, namely Src 10, Src 11 and 12
(Fig. \ref{cpart}). A background region outside the IC443 bright
thermal emission has been identified.
We have also identified the Th4 region at $\sim 1\arcmin$ from Src 12
to be used as background region for the analysis of Src 11 and 12 which
are surrounded by substantial thermal emission from the SNR.  In all
the cases the background regions have been corrected for vignetting and
area effects prior to subtraction from source spectra, using the
ancillary response files (arf) computed with the SAS task {\sc arfgen}.
As a final step, we have summed the spectra of MOS1 and MOS2, averaging
the arf files, and we have used the response matrix of MOS1 for the
analysis of the summed spectrum. The fittings have been performed on
the PN and MOS-summed spectrum for each source.

\begin{figure}
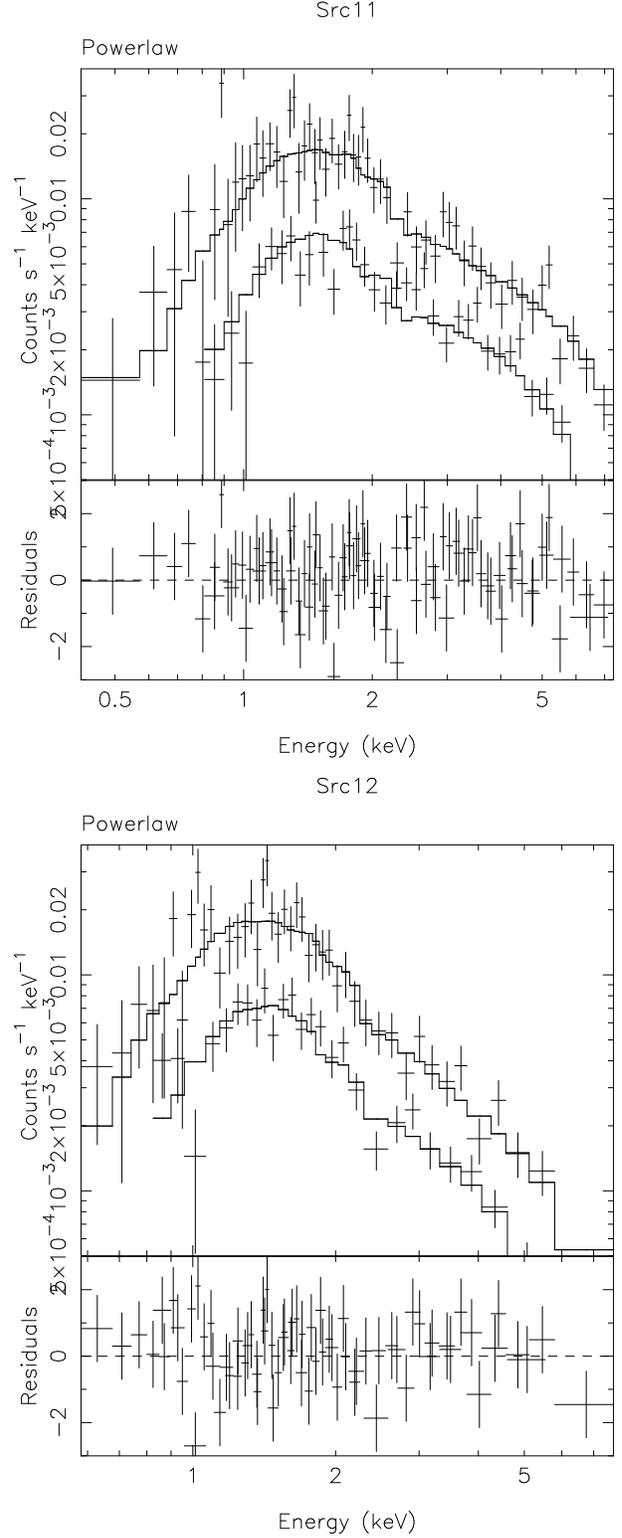

  \centerline{\psfig{file=FIGURES/h3933f5a.ps,width=8.0cm,angle=-90}}
  \medskip
  \centerline{\psfig{file=FIGURES/h3933f5b.ps,width=8.0cm,angle=-90}}
  \caption{PN+MOS spectra of Src 11 and 12 with power-law best-fit model. The best-fit parameters are reported in Table \ref{srcpar}.}
  \label{srcbsp}
\end{figure}

Src 11 and 12 were confused into a single source in a BeppoSAX
observation (\citealt{bb00}), but are clearly separated by \xmm\ into one
extended source (n.  11) plus a point-like source (n. 12, Fig.
\ref{cpart}).  Given the vicinity of these two sources to the
diffuse thermal emission, we have analyzed their spectra using the Th4
region as background. We have extracted the spectra from a circle with
radius of 16\arcsec\  and 24\arcsec\  for Src 12 and 11, respectively.
We have applied the PSF correction to the effective area used in the
fits of Src 12 spectrum only. The spectra of these two sources are
shown in Fig.  \ref{srcbsp} together with their best-fit model, while the
best-fit parameters and their uncertainties are reported in Table
\ref{srcpar}.  A fit to a combination of thermal models (not reported)
shows that the thermal nature of the spectra of these sources can be
ruled out safely.

The spectrum of Src 10 is also non-thermal. In spite of the higher
uncertainties associated with its best-fit parameters (this source is
sensibly fainter than the Src 11 and 12, see Table \ref{srcpar}), we
can say this source is characterized by a spectrum of intermediate slope
between Src 11 and 12, and a significantly lower absorption.

\begin{table}
\caption{Summary of PN spectral fitting results.}
\label{srcpar}
\medskip
\centering\begin{minipage}{8.5cm}
\begin{tabular}{lcccc} \hline
Name & \nh & $\gamma$ & flux\footnote{Absorbed power-law flux in the 0.5--10 keV band.} & $\chi^2/dof$ \\
     & $\times 10^{21}$ &  & erg cm$^{-2}$ s$^{-1}$ \\
\hline


Src10 & $1.6^{+2.2}_{-1.4}$ & $1.86^{+0.34}_{-0.49}$ & $5.0\times 10^{-14}$ & 14/17 \\
Src11 & $6.3^{+2.1}_{-1.6}$ & $1.48^{+0.20}_{-0.08}$ & $4.0\times 10^{-13}$ & 97/88 \\
Src12 & $8.1^{+2.2}_{-1.8}$ & $2.31^{+0.27}_{-0.24}$& $2.3\times 10^{-13}$ & 64/67 \\

\noalign{\smallskip}
\hline
\end{tabular}
\end{minipage}
\end{table}

\section{Discussion}
The \xmm\ observations of IC443 presented above have unprecedented
sensitivity in 2--10 keV regime. The nature of the detected localized
sources within the boundary of IC443 SNR should be a subject of a
careful inspection because some of them could be extragalactic
sources unrelated to the SNR. Using the sensitivity map of the
detection run,we have computed the log(N)-log(S) relation in 2--10 keV
band of our IC443 observations, and we have compared it with the same
function computed for the set of 10 XMM Galactic Plane Survey (GPS)
observations of \citet{bfs02}. The GPS log(N)-log(S) is systematically lower
than the relation we have found in IC443 by a factor of 2 in the range
$10^{-3}-10^{-2}$ cnt s$^{-1}$, and even more for brighter sources,
thus indicating a clear excess of bright sources in the IC443 field of
view.  It seems therefore likely that some of the detected sources are
expected to be really related to the SNR on statistical grounds.
We have therefore investigated the possible nature of compact X-ray
sources inside IC443.

\begin{figure}
  \centerline{\psfig{file=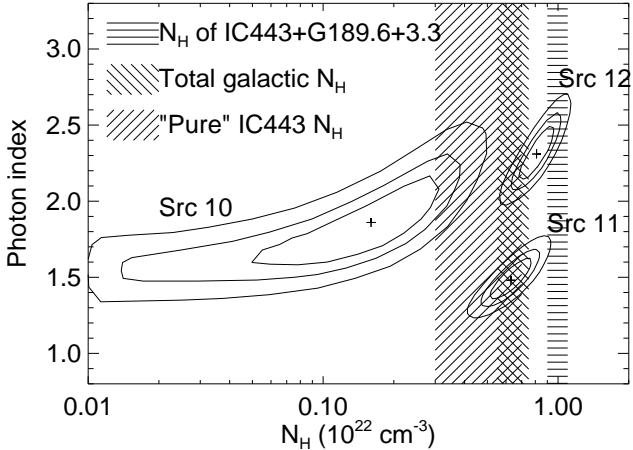,width=9.5cm}}
  \caption{Confidence level contours (68\%, 90\% and 99\%) in the \nh
  $- \gamma$ parameter space for Src 10, 11 and 12. We have also plotted
  the expected range of \nh\  for IC443 and G189.6+3.3 overlapping
  region, for IC443 only (both according \protect\citealt{aa94}), and the
  total galactic \nh\  in the direction of IC443.}

  \label{cont}
\end{figure}

The hardness ratio distributions (Fig. \ref{hr}) allow us to conclude
that only Src 7 is most likely of galactic nature, probably an X-ray
counterpart of an active star. As for the other sources, the HR diagram
does not make it possible to tell if they are galactic or
extragalactic. The results of the spectral analysis on some of the
sources give us additional informations. In Fig. \ref{cont}, we report the contour
of confidence levels for the interstellar absorption and the power-law
spectral index of Src 10, 11 and 12, along with some expected values
for \nh. We can conclude that Src 10 and 11 have the highest
probability to be of galactic origin, while for Src 12 there is a large
overlap between the allowed range of \nh\  and values well above the total
galactic \nh\  toward this direction.  However, we note that
\citet{aa94} report a total absorption of $\sim 10^{22}$ cm$^{-2}$ for
regions inside the shells of IC443 and the overlapping G189.6+3.3
SNR, as the region we are considering here. We also note that, in
addition to the absorption along the line of sight, there could be a
considerable contribution to \nh\  from the molecular cloud. For
instance, \citet{rgw95} have measured an extinction in the 2.2 $\mu
{\rm m}$ band of $\sim 1.5$ in this region, which corresponds to a value
of \nh $\sim 3\times 10^{22}$ cm$^{-2}$. Therefore, after all, there is
no strong evidence even for Src 12 to be extragalactic.

Having established that there may be a relation between the
sources and IC443, we now discuss a different scenario which
may explain the association.  There are three basic classes of
compact hard X-ray sources related to the SNR.  One of these, the
existence of a pulsar wind nebula inside IC443, has already been
discussed by \citet{ocw01} and \citet{bb01} for Src
5, and therefore we will not discuss it here.

\subsection{Shocked molecular clump}

Shocked molecular clumps could represent another class of compact hard
sources. As was discussed earlier, IC443 is known to be interacting
with a molecular cloud and the southern part of the remnant reveals a
clumped and knotty structure of the IR molecular emission down to
arcsecond scales (e.g. \citealt{rgw95}).  Recent {\it
ISO} observations (\citealt{ccp99}) of the southern region are
consistent with that expected from shocked molecular clump models.
This is also shown by Fig. \ref{cpart} which shows the optical,
infrared, and radio images of the region in which the interaction with
the cloud is occurring. In particular, the absence of bright optical
filaments is due to the very low shock velocity and very high density
confirmed by \citet{rgw95}, while the strong emission in the 2.2 $\mu
m$ image is primarily due to the 1-0 $S(1)$ line of $H_2$ at 2.122 $\mu
m$, another strong indicator of slow shock in molecular clouds. It is
remarkable that 5 out of 11 (excluding the plerion) compact X-ray
sources are located close to this infrared-filament network.

Modeling of nonthermal emission from an SNR interacting with a
molecular cloud has been recently performed by \citet{bce00}. It has been shown
that hard X-ray and \gr emission structure should consist of an
extended shell-like feature related to a radiative shock, and localized
sources corresponding to shocked molecular clumps.
\citet{bce00} predicted a hard spectrum
with a photon index below 2.0. Because of the hard spectrum and heavy
absorption, localized spots of a few arcmin size would be seen only in
hard X-rays.  The emission originates from shock-accelerated
electrons of energy below 1 GeV. The predicted flux density is $\sim 6\times
10^{-5}$ keV cm$^{-2}$ s$^{-1}$ keV$^{-1}$
at 10 keV for a 30 $\kms$ shock in a clump of
half parsec radius and density of 10$^4~ \cmc$. The predicted count rates
are consistent with the observed rate of 1SAX J0618.0+2227
(\citealt{bb00}). Recent {\it ASCA} study of  $\gamma$ Cygni,
a shell-type SNR interacting with a molecular cloud, revealed
several clumps of hard X-ray emission dominating the SNR emission above
4 keV (\citealt{uta02}). While somewhat brighter ($\sim 4.5 \times
10^{-12} \enf$) and more extended, the clumps resemble the
spectra and surface brightness of the source 1SAX J0618.0+2227
inside IC 443.

Therefore, 1SAX J0618.0+2227, located near bright spots of
molecular hydrogen emission, could be a shocked molecular clump.
However, with the \xmm\ PN camera the source  1SAX J0618.0+2227 is
resolved into two different sources, Src 11 and 12 (Fig.
\ref{hardpn}). Src 12 apparently has a soft spectrum and a size of
12\arcsec\  FWHM, which makes the clump interpretation
unlikely. The Src 11 has a hard spectrum, but still only
a size of 20\arcsec\  FWHM ($\sim 4\times 10^{17}$ cm at 1.5 kpc). The
shocked molecular clump interpretation of Src 11 would require a
substantially higher shock velocity in the clump than  30 $\kms$
to match the observed flux. The energetic requirements are rather
stringent for such a small scale shock.

\subsection{Fast-moving isolated ejecta fragments}

The third class of possible compact hard X-ray sources related to
the SNR are the isolated ejecta fragments.
Multiwavelength studies of the structure of supernova remnants have revealed
the presence of fine-scale structure with rich emission spectra.  There
are metal-rich fragments of SN ejecta interacting with surrounding
media as well as circumstellar mass-loss matter.
Some X-ray knots, e.g. the silicon-rich shrapnel, were
observed in the Vela SNR by \citet{aet95} and
\citet{mta01} and also
in the Tycho SNR with \xmm\ by \citet{dsa01}.
Supersonic motion of the knots in the ambient medium will result in a bow
shock/knot-shock structure creation.
It was shown that an individual knot propagating through
a molecular cloud would be observable with \xmm\
and \chan\ from a few kpc distance (\citealt{byk02}).
Such a know would have a hard X-ray spectrum of
photon index $\lsim 1.5$ in 1--10 keV regime,
with prominent lines of some metals e.g. Si, S, Ar, Ca, Fe etc.

\begin{figure}
  \centerline{\psfig{file=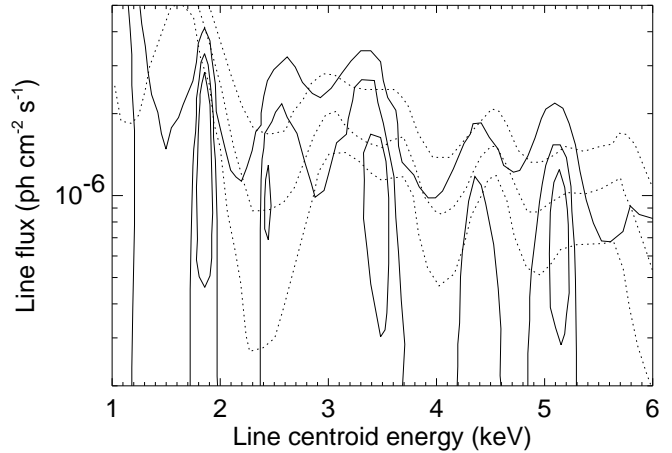,width=9.5cm}}
  \caption{Confidence level contours (68\%, 90\% and 99\%) in the line
  centroid vs. line flux parameter space for Src 11 (solid lines) and 12 (dotted
  lines). We have used a model with a single gaussian line of variable width
  plus a power-law continuum.  There is no firm detection
  of any line; the 99\% contour levels (the uppermost for each source)
  define our estimate of the upper-limit to the line flux.}

  \label{line}
\end{figure}

In order to verify the knot hypothesis for Src 11 and Src 12, we
investigated the presence of emission lines in the X-ray
spectrum. We added a gaussian line of variable width in the
range 0--200 eV to the power-law model and we fitted the spectra,
thus deriving the $\chi^2$ contour levels for the line centroid
and flux shown in Fig. \ref{line}. We did not detect any
emission lines at the 99\% confidence level, even though there is
excess flux at $\sim$ 1.8 keV,
3.4--3.6 keV and 5.1 keV only for Src 11. Fig.
\ref{line} reports the upper-limit to the line flux at
different centroid energies. We have verified that the excess flux
cannot be explained by known background emission lines.

We applied the supersonic knot model of \citet{byk02} to Src 11.
X-ray line emission was simulated from a metal rich ejecta fragment of
$3\times 10^{17}$ cm radius ($\sim 15\arcsec$ at 1.5 kpc) and a
mass of $\sim 10^{-2} \Msun$ interacting with a molecular cloud.
The simulated fragment mass was dominated by
oxygen with a minor mass contribution ($\sim 10^{-4} \Msun$) of
other SN nucleosynthesis products like Si, S, Ar, Ca, Fe. The knot
travels through an inter-clump medium of number density $\lsim
100 \cmc$ at a velocity of $\sim 500 \kms$. Under these
conditions, the model yields a 1.8 keV Si line flux of $\gsim
10^{-6}$ ph cm$^{-2}$ s$^{-1}$ (if the electron diffusion
coefficient is $\lsim 10^{18}$ cm$^2$ s$^{-1}$), to be compared
with our upper-limit of $4\times 10^{-6}$ ph cm$^{-2}$ s$^{-1}$.
The optical depth of the 1.8 keV Si line (Si VI and higher
ionization states) is $\sim 6$ in that case, implying a
saturation effect, i.e. increasing the Si mass would not
increase the observed flux, if the knot radius is the same.  The
depth is somewhat lower for heavier elements where the saturation
becomes improtant only above $10^{-4} \Msun$.  A higher knot velocity
(or lower diffusion coefficients) would produce higher Si line
fluxes, as so would a bigger knot. However, the apparent
size of Src 11 would point against knots having sizes sensibly
different from those used in our simulation. The knot position, given
the SNR age above 10 kyr, is constraining  the knot velocity. Thus
the electron diffusion coefficient is the most important parameter,
which may be varied to match more accurate measurement of the 1.8
Si emission line flux.

We note that the lifetime of the ejecta fragment in the
inter-clump medium is rather short, being below 100 yr. This
implies that the fragment should spent most of its time in a tenuous
gas of density $\lsim 0.1 \cmc$ of the SNR interior to survive
during the lifetime of IC443, which is expected to be about 30,000 yr (e.g.
\citealt{che99}).  Indeed, the Src 11 is located just on the border
of the molecular cloud, which suggests that the interaction with
denser material has just started. Note that the [Ar
III] bright knots of velocities $\sim 500 \kms$ (but with a somewhat
smaller size) observed in the Crab nebula with {\it HST} by
\citet{sf02} could represent a similar phenomenon

The definite test to distinguish between the two interpretations, namely
a fast moving metal-rich knot or a shocked clump, would be an X-ray spectral
lines study because the shocked molecular clump emission would be dominated
by a hard continuum in contrast to the metal-rich knot case.
High-quality line spectra of the Src 11 would allow the X-ray study of
SN nucleosynthesis.

\section{Summary and conclusions}

We have detected 12 compact hard X-ray sources in the field of the
large SNR IC443, more than the expected number of galactic and extragalactic
sources in this field, thus suggesting an association with the SNR for some
of them. We have carried out a spectral analysis of the sources,
by fitting the X-ray spectra of the brightest ones, and by
computing X-ray hardness ratios. We have found that all
sources but one have a clearly non-thermal spectrum. We have noted
that 6 (including the plerion) out of 12 sources are concentrated
in a small ($\sim 15\arcmin$) area of the SNR, which was
previously known to be a region of interaction between the SNR
shock and a molecular cloud. We suggest that some of the sources
may indeed be related to the SNR itself, and we review the
possible formation mechanisms of formation of isolated hard X-ray sources in
SNR shell. We have considered both electron acceleration in
molecular clump shocks, and the emission from ejecta fragment as
an alternative model. The latter is suggested by a modest
evidence of a Si line in the spectrum of one of the sources. A
deeper X-ray spectroscopy observation of the sources in IC443 and
more X-ray observations of other SNRs interacting with molecular
clouds are urgently needed to shed more light on this intriguing
new class of hard X-ray sources.

\begin{acknowledgements}
F.B. acknowledges support from European Space Agency, Ministero
dell'Universit\`a e della Ricerca Scientifica, and Agenzia Spaziale
Italiana. The work of A.M.B was supported by
INTAS-ESA 99-1627 grant.

\end{acknowledgements}
\appendix

\bibliographystyle{aa}
\bibliography{references}

\end{document}